%% file: covidnet.tex
\documentclass[fleqn,10pt]{wlscirep}
\usepackage[utf8]{inputenc}
\usepackage[T1]{fontenc}
\usepackage{adjustbox}

\title{COVID-Net: A Tailored Deep Convolutional Neural Network Design for Detection of COVID-19 Cases from Chest X-Ray Images}

\author[1,2,3*]{Linda Wang}
\author[1,2,3]{Zhong Qiu Lin}
\author[1,2,3]{Alexander Wong}
\affil[1]{Department of Systems Design Engineering, University of Waterloo, Canada}
\affil[2]{Waterloo Artificial Intelligence Institute, Canada}
\affil[3]{DarwinAI Corp., Canada}

\affil[*]{linda.wang@uwaterloo.ca}

\begin{abstract}
The COVID-19 pandemic continues to have a devastating effect on the health and well-being of the global population.  A critical step in the fight against COVID-19 is effective screening of infected patients, with one of the key screening approaches being radiology examination using chest radiography.  It was found in early studies that patients present abnormalities in chest radiography images that are characteristic of those infected with COVID-19.  Motivated by this and inspired by the open source efforts of the research community, in this study we introduce COVID-Net, a deep convolutional neural network design tailored for the detection of COVID-19 cases from chest X-ray (CXR) images that is open source and available to the general public. To the best of the authors' knowledge, COVID-Net is one of the first open source network designs for COVID-19 detection from CXR images at the time of initial release.  We also introduce COVIDx, an open access benchmark dataset that we generated comprising of 13,975 CXR images across 13,870 patient patient cases, with the largest number of publicly available COVID-19 positive cases to the best of the authors' knowledge.  Furthermore, we investigate how COVID-Net makes predictions using an explainability method in an attempt to not only gain deeper insights into critical factors associated with COVID cases, which can aid clinicians in improved screening, but also audit COVID-Net in a responsible and transparent manner to validate that it is making decisions based on relevant information from the CXR images.  By no means a production-ready solution, the hope is that the open access COVID-Net, along with the description on constructing the open source COVIDx dataset, will be leveraged and build upon by both researchers and citizen data scientists alike to accelerate the development of highly accurate yet practical deep learning solutions for detecting COVID-19 cases and accelerate treatment of those who need it the most. 
\end{abstract}
\begin{document}

\flushbottom
\maketitle

\thispagestyle{empty}

\section{Introduction}
\input{content/introduction.tex}

\section{Methods}
\input{content/methods}

\section{Experimental Results}
\input{content/results}

\section{Conclusion}
\input{content/conclusion}

\section*{Acknowledgements}

We would like to thank Natural Sciences and Engineering Research Council of Canada (NSERC), the Canada Research Chairs program, DarwinAI Corp., and Audrey Chung.  

\section*{Author contributions statement}
L.W. and A.W. conceived the experiment,  L.W., Z.L. and A.W. conducted the experiment, L.W. and A.W. analysed the results.  All authors reviewed the manuscript. 

\section*{Additional information}

\textbf{Competing interests} L.W., Z.L. and A.W. are affiliated with DarwinAI Corp. 

{\small
\bibliography{egbib}
}
\end{document}

%% file: content/introduction.tex
The COVID-19 pandemic continues to have a devastating effect on the health and well-being of the global population, caused by the infection of individuals by the severe acute respiratory syndrome coronavirus 2 (SARS-CoV-2).  A critical step in the fight against COVID-19 is effective screening of infected patients, such that those infected can receive immediate treatment and care, as well as be isolated to mitigate the spread of the virus.  The main screening method used for detecting COVID-19 cases is reverse transcriptase-polymerase chain reaction (RT-PCR)~\cite{PCR} testing, which can detect SARS-CoV-2 RNA from respiratory specimens (collected through a variety of means such as nasopharyngeal or oropharyngeal swabs). While RT-PCR testing is the gold standard as it is highly specific, it is a very time-consuming, laborious, and complicated manual process that is in short supply. Furthermore, the sensitivity of RT-PCR testing is highly variable and have not been reported in a clear
and consistent manner to date~\cite{West}, and initial findings in China showing relatively poor sensitivity~\cite{Fang}.  Furthermore, subsequent findings showed highly variable positive rate depending on how the specimen was collected as well as decreasing positive rate with time after symptom onset~\cite{Yang2020.02.11.20021493,Wikramaratna2020.04.05.20053355}. 

An alternative screening method that has also been utilized for COVID-19 screening has been radiography examination, where chest radiography imaging (e.g., chest X-ray (CXR) or computed tomography (CT) imaging) is conducted and analyzed by radiologists to look for visual indicators associated with SARS-CoV-2 viral infection.  It was found in early studies that patients present abnormalities in chest radiography images that are characteristic of those infected with COVID-19~\cite{Ng,Huang,Guan}, with some suggesting that radiography examination could be used as a primary tool for COVID-19 screening in epidemic areas~\cite{Ai}.  For example, Huang et al.~\cite{Huang} identified that the majority of the COVID-19 positive cases in their study presented bilateral radiographic abnormalities in CXR images, and Guan et al.~\cite{Guan} identified COVID-19 positive cases in their study presented radiographic abnormalities such as ground-glass opacity, bilateral abnormalities, and interstitial abnormalities in CXR and CT images.  While much of the recent discussion has revolved around CT imaging due to greater image detail in the acquisitions, there are several advantages to leveraging CXR imaging for COVID-19 screening amid the global COVID-19 pandemic, particularly in resource-constrained areas and heavily-affected areas:
\begin{itemize}
\item \textbf{Rapid triaging}: CXR imaging enables rapid triaging of patients suspected of COVID-19 and can be done in parallel of viral testing (which takes time) to help relief the high volumes of patients especially in areas most affected where they have ran out of capacity (e.g., New York, Spain, and Italy), or even as standalone when viral testing isn't an option (low supplies).  Furthermore, CXR imaging can be quite effective for triaging in geographic areas where patients are instructed to stay home until the onset of advanced symptoms (e.g., New York City), since abnormalities are often seen at time of presentation when patients suspected of COVID-19 arrive at clinical sites~\cite{RSNA}. 
\item \textbf{Availability and Accessibility}: CXR imaging is readily available and accessible in many clinical sites and imaging centers as it is considered standard equipment in most healthcare systems.  In particular, CXR imaging is much more readily available than CT imaging, especially in developing countries where CT scanners are cost prohibitive due to high equipment and maintenance costs.
\item \textbf{Portability}: The existence of portable CXR systems means that imaging can be performed within an isolation room, thus significantly reducing the risk of COVID-19 transmission during transport to fixed systems such as CT scanners as well as within the rooms housing the fixed imaging systems~\cite{RSNA}.  
\end{itemize}

\begin{figure}[t]\center
  \includegraphics[width=0.4\textwidth]{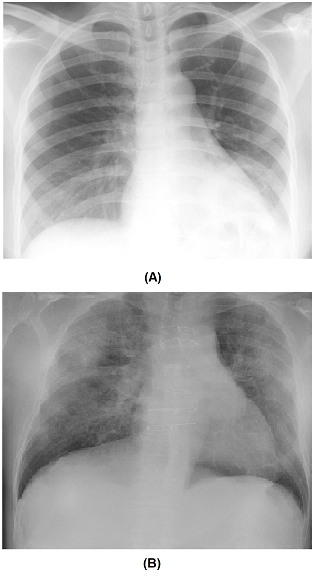}
  \caption{Example CXR images of: (A) non-COVID19 infection, and (B) COVID-19 viral infection in the COVIDx dataset.}
  \label{fig1}
  \vspace{-15pt}
\end{figure}

As such, radiography examination can be conducted faster and have greater availability given the prevalence of chest radiology imaging systems in modern healthcare systems and the availability of portable units, making them a good complement to RT-PCR testing particularly since CXR imaging is often performed as part of standard procedure for patients with a respiratory complaint~\cite{BSTI}.  Furthermore, some have suggested that as the COVID-19 pandemic progresses, there will be a greater reliance on portable CXR due to the aforementioned advantages~\cite{Jacobi}.  However, one of the biggest bottlenecks faced is the need for expert radiologists to interpret the radiography images, since the visual indicators can be subtle.  As such, computer-aided diagnostic systems that can aid radiologists to more rapidly and accurately interpret radiography images to detect COVID-19 cases is highly desired.  

Motivated by the urgent need to develop solutions to aid in the fight against the COVID-19 pandemic, inspired by the open source and open access efforts of the research community, and intrigued in exploring the efficacy of AI systems leveraging the more readily available and accessible CXR imaging modality, this study introduces COVID-Net, a deep convolutional neural network design tailored for the detection of COVID-19 cases from CXR images that is open source and available to the general public.  To the best of the authors' knowledge, COVID-Net is one of the first open source network designs for COVID-19 detection from CXR images at the time of initial release.  We also introduce COVIDx, an open access benchmark dataset that we generated comprising of 13,975 CXR images across 13,870 patient cases, created as a combination and modification of five open access data repositories containing chest radiography images (i.e.,~\cite{Cohen},~\cite{Figure1},~\cite{actualmed},~\cite{kaggle2}, and~\cite{kaggle}), two of which we introduced~\cite{Figure1,actualmed}.  To the best of the authors' knowledge, COVIDx is the largest open access benchmark dataset available in terms of the number of publicly available COVID-19 positive cases.  Furthermore, we investigate how COVID-Net makes predictions using an explainability method in an attempt to not only gain deeper insights into critical factors associated with COVID cases, which can aid clinicians in improved screening, but also audit COVID-Net in a responsible and transparent manner to validate that it is making decisions based on relevant information from the CXR images.    

The paper is organized as follows.  First, Section 2 discusses the strategy used to create the COVIDx dataset, the strategy leveraged to create the proposed COVID-Net, the architecture design of COVID-Net, the implementation details of COVID-Net, and the strategy leveraged to audit COVID-Net via explainability.  Section 3 presents and discusses the results of experiments conducted to evaluate the efficacy of the proposed COVID-Net in both a quantitative and qualitative manner.  Finally, conclusions are drawn and future directions discussed in Section 4. 

\section{Related Work}

Motivated by the need for faster interpretation of radiography images, a number of artificial intelligence (AI) systems based on deep learning~\cite{Deeplearning} have been proposed and results have shown to be quite promising in terms of accuracy in detecting patients infected with COVID-19 via radiography imaging, with the focus primarily on CT imaging~\cite{Goze,Xu,Li,shi2020largescale}.  However, to the best of the authors' knowledge at the time of the initial release of the proposed COVID-Net, most of the developed AI systems proposed in research literature have been closed source and unavailable to the research community to build upon for deeper understanding and extension of these systems. Furthermore, most of these systems are unavailable for public access and use.  There has been significant recent efforts to push for open access and open source AI solutions for radiography-driven COVID-19 case detection~\cite{Cohen,Figure1,actualmed}, with an exemplary effort being the open source COVID-19 Image Data Collection, an effort by Cohen et al.~\cite{Cohen} to build a dataset consisting of COVID-19 cases (as well as SARS and MERS cases) with annotated CXR and CT images, so that the research community and citizen data scientists can leverage the dataset to explore and build AI systems for COVID-19 detection.

Since the time of the initial public release of the proposed COVIDx dataset and the proposed COVID-Net, a number of studies have been conducted in the area of COVID-19 detection using CXR images~\cite{li2020covidmobilexpert, minaee2020deepcovid, afshar2020covidcaps, luz2020effective, Khobahi2020.04.14.20065722,ucar,tartaglione2020unveiling,yeh2020cascaded,zhang2020covidda}, several of which have leveraged variants of the COVIDx dataset or COVID-Net to conduct such studies~\cite{afshar2020covidcaps, luz2020effective, Khobahi2020.04.14.20065722,ucar,tartaglione2020unveiling,yeh2020cascaded,zhang2020covidda}.  The majority of such studies have focused on the exploration of deep convolutional neural networks, given the significant successes and state-of-the-art achieved in a variety of computer vision tasks.  It is important to note that the proposed COVID-Net and corresponding COVIDx dataset continues to evolve as new patient cases are continuously added and are made available publicly on a regular basis, and thus this study represents a snapshot of the current state of COVID-Net and COVIDx. 

%% file: content/methods.tex
\begin{figure}[t]
\centering
  \includegraphics[width=0.5\linewidth]{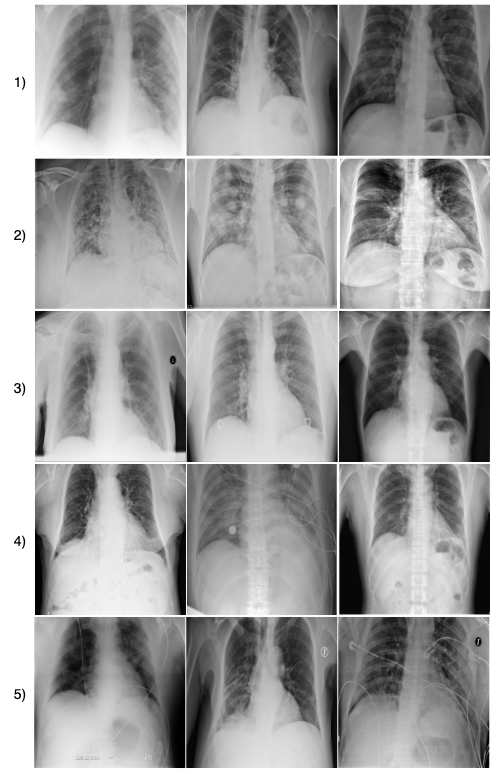}
  \caption{Example CXR images from the COVIDx dataset, which comprises of 13,975 CXR images across 13,870 patient cases from five open access data repositories: 1. COVID-19 Image Data Collection~\cite{Cohen}, 2. Figure 1 COVID-19 Chest X-Ray Dataset Initiative~\cite{Figure1} (established with Figure 1), 3. RSNA Pneumonia Detection challenge dataset~\cite{kaggle}, 4. ActualMed COVID-19 Chest X-Ray Dataset Initiative~\cite{actualmed} (established with ActualMed), and 5. COVID-19 radiography database~\cite{kaggle2}. }
  \label{fig:covidx}
\end{figure}

In this study, a human-machine collaborative design strategy is leveraged to create COVID-Net, where human-driven principled network design prototyping is combined with machine-driven design exploration to produce a network architecture tailored for the detection of COVID-19 cases from CXR images.  An open access benchmark dataset called COVIDx is also created to facilitate for training and evaluating COVID-Net.  Here, we will discuss in detail the process of creating the COVIDx dataset, the architecture design methodology behind the proposed COVID-Net, the resulting network architecture, the implementation details in creating COVID-Net, and the strategy for auditing COVID-Net via explainability.  

\begin{figure*}[h]
\centering
  \includegraphics[width=0.7\linewidth]{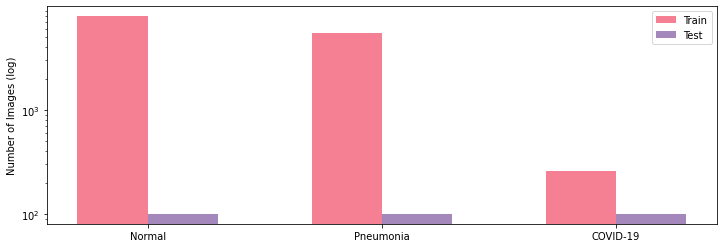}
  \caption{CXR images distribution for each infection type of the COVIDx dataset (normal means no infection). (Left bar) number of training images, (right bar) number of test images.}
  \label{fig:images-dist}
\end{figure*}

\begin{figure*}[h]
    \centering
    \includegraphics[width=0.7\linewidth]{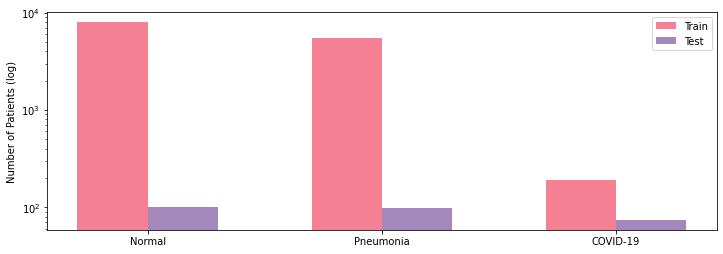}
    \caption{Number of patient cases for each infection type of the COVIDx dataset (normal means no infection). (Left bar) number of patient cases for training, (right bar) number of patient cases for testing.}
    \label{fig:patients-dist}
\end{figure*}

\subsection{COVIDx Dataset}

The dataset used to train and evaluate the proposed COVID-Net, which we will refer to as COVIDx, is comprised of a total of 13,975 CXR images across 13,870 patient cases.  To the best of the authors' knowledge, the proposed COVIDx dataset is the largest open access benchmark dataset in terms of the number of COVID-19 positive patient cases.  To generate the COVIDx dataset, we combined and modified five different publicly available data repositories: 1) COVID-19 Image Data Collection~\cite{Cohen}, 2) Figure 1 COVID-19 Chest X-ray Dataset Initiative~\cite{Figure1}, established in collaboration with Figure 1, 3) ActualMed COVID-19 Chest X-ray Dataset Initiative~\cite{actualmed}, established in collaboration with ActualMed, 4) RSNA Pneumonia Detection Challenge dataset~\cite{kaggle}, which used publicly available CXR data from \cite{wang2017chestxray}, and 5) COVID-19 radiography database~\cite{kaggle2}.  Example CXR images from the COVIDx dataset are shown in Fig.~\ref{fig:covidx} in illustrate the diversity of patient cases in the dataset.  The choice of these five datasets from which to create COVIDx is guided by the fact that all five of the datasets are open source and fully accessible to the research community and the general public, and as datasets grow we will continue to grow COVIDx accordingly.  

More specifically, we combined and modified the five data repositories to create the COVIDx dataset by leveraging the following types of patient cases from each of the data repositories: 
\begin{itemize}
\item Non-COVID19 pneumonia patient cases and COVID-19 patient cases from the COVID-19 Image Data Collection~\cite{Cohen}
\item COVID-19 patient cases from the Figure 1 COVID-19 Chest X-ray Dataset Initiative~\cite{Figure1},
\item COVID-19 patient cases from the ActualMed COVID-19 Chest X-ray Dataset Initiative~\cite{actualmed}
\item Patient cases who have no pneumonia (i.e., normal) and non-COVID19 pneumonia patient cases from RSNA Pneumonia Detection Challenge dataset~\cite{kaggle} 
\item COVID-19 patient cases from COVID-19 radiography database~\cite{kaggle2}
\end{itemize}  The distribution of images and patient cases amongst the different infection types shown in Fig.~\ref{fig:images-dist} and \ref{fig:patients-dist}, respectively. The most noticeable trend is the limited amount of COVID-19 infection cases and associated CXR images, which reflects the scarcity of COVID-19 case data available in the public domain but also highlights the need to obtain more COVID-19 data as more case data becomes available to improve the dataset. More specifically, the COVIDx dataset contains  358 CXR images from 266 COVID-19 patient cases. For CXR images with no pneumonia and non-COVID19 pneumonia, there are significantly more patient cases and corresponding CXR images. More specifically, there are a total of 8,066 patient cases who have no pneumonia (i.e., normal) and 5,538 patient cases who have non-COVID19 pneumonia.  Dataset generation scripts for constructing the COVIDx dataset is available publicly for open access at \url{https://github.com/lindawangg/COVID-Net}.

\subsection{Principled network design prototyping} The first stage of the human-machine collaborative design strategy employed to create the proposed COVID-Net is a principled network design prototyping stage, where an initial network design prototype is constructed based on human-driven design principles and best practices. More specifically in this study, we leveraged residual architecture design principles \cite{resnet} as they have been shown time and again to enable reliable neural network architectures that are easier to train to high performance, and enables deeper architectures to be built successfully.  

In this study, we construct the initial network design prototype to make one of the following three predictions: a)  no infection (normal), b) non-COVID19 infection (e.g., viral, bacterial, etc.), and c) COVID-19 viral infection (see Fig.~\ref{fig1} for example CXR images of non-COVID19 and COVID-19 infections).  The rationale for choosing these three possible predictions is that it can aid clinicians to better decide not only who should be prioritized for PCR testing for COVID-19 case confirmation, but also which treatment strategy to employ depending on the cause of infection, since COVID-19 and non-COVID19 infections require different treatment plans. 

\begin{figure*}[ht]
\centering
  \includegraphics[width=1\textwidth]{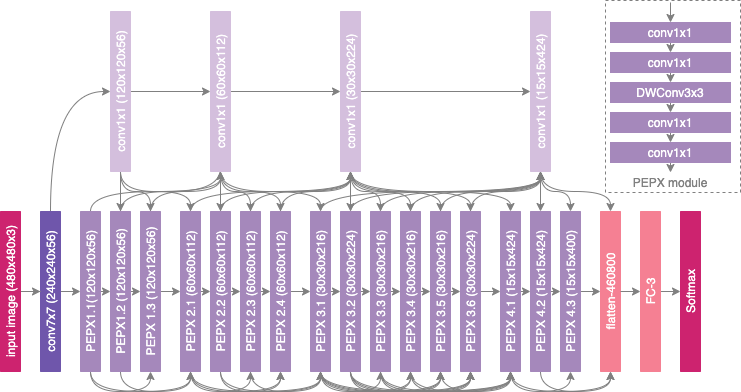}
  \caption{\textbf{COVID-Net Architecture.} High architectural diversity and selective long-range connectivity can be observed as it is tailored for COVID-19 case detection from CXR images. The heavy use of a projection-expansion-projection design pattern in the COVID-Net architecture can also be observed, which provides enhanced representational capacity while maintaining computational efficiency.}
  \label{fig:arch}
\end{figure*}

\subsection{Machine-driven Design Exploration} 
\label{sec:gensynth}
The second stage of the human-machine collaborative design strategy employed to create the proposed COVID-Net is a machine-driven design exploration stage.  More specifically, at this stage, the initial network design prototype, data, along with human specific design requirements, act as a guide to a design exploration strategy to learn and identify the optimal macroarchitecture and microarchitecture designs with which to construct the final tailor-made deep neural network architecture. Such a machine-driven design exploration stage enables much greater granularity and much greater flexibility than is possible through manual human-driven architecture design, while still ensuring that the resulting deep neural network architecture satisfies domain-specific operational requirements.  This is especially important for the design of COVID-Net, where sensitivity to COVID-19 cases is significant to limit the number of missed COVID-19 cases as much as possible.

In this study, we leverage generative synthesis~\cite{wong2018ferminets} as the machine-driven design exploration strategy, which is based on an intricate interplay between a generator-inquisitor pair that work in tandem to garner insights and learn to generate deep neural network architectures that best satisfies human specified design requirements.  More specifically, the following human specified design requirements were employed in this study to enable the generative synthesis process to learn and identify the optimal macroarchitecture and microarchitecture designs for the final COVID-Net network architecture: (i) COVID-19 sensitivity $\geq$ 80\%, and (ii) COVID-19 positive predictive value (PPV) $\geq$ 80\%. By employing the aforementioned human specified design requirements, the machine-driven design exploration can be conducted in a way that ensures that the resulting COVID-Net is able to detect COVID-19 positive cases while limiting the number of false positive COVID-19 detections to avoid overwhelming clinical sites with unnecessary burden.

\subsection{COVID-Net Network Architecture} The proposed COVID-Net network architecture is shown in Fig.~\ref{fig:arch}, and available publicly for open access at \url{https://github.com/lindawangg/COVID-Net}. A number of interesting observations can be made with regards to the COVID-Net network arhchitecture design that was created via a human-machine collaborative design strategy.

\subsubsection{Lightweight design pattern} It can be observed that the COVID-Net network architecture makes heavy use of a lightweight residual projection-expansion-projection-extension (PEPX) design pattern, which consists of:

\begin{itemize}
    \item \textbf{First-stage Projection:} 1$\times$1 convolutions for projecting input features to a lower dimension, 
    \item \textbf{Expansion:} 1$\times$1 convolutions for expanding features to a higher dimension that is different than that of the input features, \item \textbf{Depth-wise Representation:} efficient 3$\times$3 depth-wise convolutions for learning spatial characteristics to minimize computational complexity while preserving representational capacity, 
    \item \textbf{Second-stage Projection:} 1$\times$1 convolutions for projecting  features back to a lower dimension, and
    \item \textbf{Extension:} 1$\times$1 convolutions that finally extend channel dimensionality to a higher dimension to produce the final features.
    \end{itemize}
Such a customized lightweight design pattern, which is discovered by the machine-driven design exploration strategy and not previously introduced in literature to the best of the authors' knowledge, enables enhanced representational capacity while maintaining reduced computational complexity.

\subsubsection{Selective long-range connectivity} 
It can also be observed that the COVID-Net architecture possesses selective long-range connectivity at various areas of the network architecture.  The use of long-range connectivity has been found previously to enable improved representational capacity as well as make it easier to train, with an exemplary network architecture leveraging long-range connectivity being densely-connected deep neural network architectures~\cite{densenet}.  However, a disadvantage of leveraging a large number of long-range connections as done in densely-connected deep neural network architectures is a noticeable increase in computational complexity as well as memory overhead. To alleviate this, it can be observed that the COVID-Net network architecture only leverages long range connections in a sparing manner where necessary, as exhibited by the existence of four densely connected $1\times1$ convolution layers that act as central hubs of long-range connectivity for earlier layers to connected to much later layers in the network architecture.  As such, the COVID-Net network architecture is able to achieve high  representational capacity and improve ease of training while still maintaining computational and memory efficiency.  This also highlights the advantages of a machine-driven design strategy for identifying the optimal connectivity within a deep neural network architecture, tailored in this case specifically for the task of COVID-19 detection based on CXR images.

\subsubsection{Architectural diversity}
Finally, it can be observed that there is considerable 
architectural diversity in the COVID-Net architecture.  In particular, the COVID-Net network architecture is comprised of a heterogeneous mix of convolution layers with a diversity of kernel sizes (ranging from $7\times7$ to $1\times1$), and grouping configurations (ranging from ungrouped to depth-wise).  The considerable architectural diversity exhibited by the COVID-Net architecture further reinforces the fact that the machine-driven design exploration strategy has tailored the network architecture at a very fine level of granularity for COVID-19 case detection from CXR images to achieve strong representational capacity for a specific task.

\subsection{Implementation Details}
The proposed COVID-Net was pretrained on the ImageNet~\cite{deng2009imagenet} dataset and then trained on the COVIDx dataset using the Adam optimizer using a learning rate policy where the learning rate decreases when learning stagnates for a period of time (i.e., 'patience').  The following hyperparameters were used for training: learning rate=2e-4, number of epochs=22, batch size=64, factor=0.7, patience=5. Furthermore, data augmentation was leveraged with the following augmentation types: translation, rotation, horizontal flip, zoom, and intensity shift.  Finally, we introduce a batch re-balancing strategy to promote better distribution of each infection type at a batch level.  The initial COVID-Net prototype was built and evaluated using the Keras deep learning library with a TensorFlow backend. The proposed COVID-Net architecture was built using generative synthesis~\cite{wong2018ferminets}, as described in Section \ref{sec:gensynth}.

\subsection{COVID-Net Auditing via Explainability}

Due to the mission-critical nature of clinical applications such as COVID-19 detection that can affect the health and well-being of patients, it is important to design deep neural network architectures such as COVID-Net with responsibility and transparency in mind.  Therefore, in this study, we perform an explainability-driven audit on COVID-Net to validate that it is making detection decisions based on relevant information rather than improper information (e.g., erroneous visual indicators outside of the body, embedded markup symbols, imaging artifacts, etc.).  More specifically, we audit COVID-Net via an qualitative analysis to study the critical factors leveraged by COVID-Net in making detection decisions.  Here, we leveraged GSInquire~\cite{lin2019explaining}, an explainability method that is a critical aspect of the generative synthesis strategy~\cite{wong2018ferminets} leveraged in the machine-driven exploration strategy used to create the proposed COVID-Net network architecture.  A brief summary of the GSInquire is provided as follows.  

GSInquire revolves around the notion of an inquisitor $\mathcal{I}$ within a generator-inquisitor pair $\left\{\mathcal{G},\mathcal{I}\right\}$, with $\mathcal{G}$ denoting a generator, that work in tandem to obtain improved insights about deep neural networks as well as learn to generate networks.  The insights gained by $\mathcal{I}$ can not only be used to improve $\mathcal{G}$ to generate better networks, but also be subsequently transformed into an interpretation of decisions made by a network.  More specifically, a deep neural network is defined as a graph $N=\left\{V,E\right\}$, comprising a set $V$ of vertices $v \in V$ and a set $E$ of edges $e \in E$ that form the network.  A generator function is defined as $\mathcal{G}(s;\theta_\mathcal{G})$ parameterized by $\theta_\mathcal{G}$ that, given a seed $s \in S$, generates a deep neural network $N_s=\left\{V_s,E_s\right\}$ (i.e., $N_s = \mathcal{G}(s)$), where $S$ is the set of possible seeds.  Finally, an inquisitor function is defined as $\mathcal{I}(\mathcal{G};\theta_\mathcal{I})$  parameterized by $\theta_\mathcal{I}$ that, given a generator $\mathcal{G}$, produces a set of parameter changes $\Delta\theta_\mathcal{G}$ (i.e., $\Delta\theta_\mathcal{G} = \mathcal{I}(\mathcal{G})$).

In the scenario where the underlying goal is to obtain an interpretation $z$ of a decision made by a reference network $N_{ref}$ (in this case, COVID-Net) for an input signal $x$ (in this case, a CXR image), both $\theta_{\mathcal{G}}$ and $\theta_{\mathcal{I}}$ are initialized based on $\left\{V_{ref},E_{ref}\right\}$, a universal performance function $\mathcal{U}$ (e.g.,~\cite{wong2019netscore}), and an indicator function $1_r(\cdot)$ s.t. $\left\{V_{s},E_{s}\right\}=\left\{V_{ref},E_{ref}\right\}$ to ensure interpretation consistency for $N_{ref}$.  
Given the generated $N_s = \mathcal{G}(s)$, the inquisitor $\mathcal{I}$ probes $\left\{\mathcal{V}_{s},\mathcal{E}_{s}\right\}$, where 
$\mathcal{V}_{s} \subseteq V_{s}$ and $\mathcal{E}_{s} \subseteq E_{s}$, with the targeted stimulus signal as $x$ 
and the corresponding set $Y_{\mathcal{G}\left(s\right)}$ of reactionary response signals 
$y \in Y_{\mathcal{G}\left(s\right)}$ are observed.  
The parameters $\theta_{\mathcal{I}}$ are updated based on $Y_{\mathcal{G}\left(s\right)}$, $\mathcal{U}(\mathcal{G}\left(s\right))$, and $1_r(\mathcal{G}\left(s\right))$, leading to the inquisitor $\mathcal{I}$ learning from the insights that are derived from $Y_{\mathcal{G}\left(s\right)}$.  Following the update of $\theta_{\mathcal{I}}$, set of parameters $\Delta\theta_\mathcal{G} = \mathcal{I}(\mathcal{G})$ is generated which can not only be leveraged to update $\theta_{\mathcal{G}}$  to improve $\mathcal{G}$, but can also be transformed and projected into same subspace as $x$ via a transformation  $\mathcal{T}(\Delta\theta_{\mathcal{G}\left(s\right)})$ to produce an interpretation $z(x;N_{ref})$.  In this study, the produced interpretation indicates the critical factors leveraged by COVID-Net in making a detection decision based on a CXR image, and can be visualized spatially relative to the CXR image for greater insights into whether COVID-Net is making the right decisions for the right reasons and validate its performance.

%% file: content/results.tex
To evaluate the efficacy of the proposed COVID-Net, we perform both quantitative and qualitative analysis to get a better understanding of its detection performance and decision-making behaviour.  

\subsection{Quantitative Analysis} To investigate the proposed COVID-Net in a quantitative manner, we computed the test accuracy, as well as sensitivity and positive predictive value (PPV) for each infection type, on the aforementioned COVIDx dataset.  The test accuracy, along with the architectural complexity (in terms of number of parameters) and computational complexity (in terms of number of multiply-accumulation (MAC) operations) are shown in Table \ref{tab:static_comp_table}.  It can be observed that COVID-Net achieves good accuracy by achieving 93.3\% test accuracy, thus highlighting the efficacy of leveraging a human-machine collaborative design strategy for creating highly-customized deep neural network architectures in an accelerated manner, tailored around task, data, and operational requirements.  This is especially important for scenarios such as disease detection, where new cases and new data are collected continuously and the ability to rapidly generate new deep neural network architectures tailored to the ever-evolving knowledge base over time is highly desired.  

\begin{figure}[h]\center
  \includegraphics[width=0.5\linewidth]{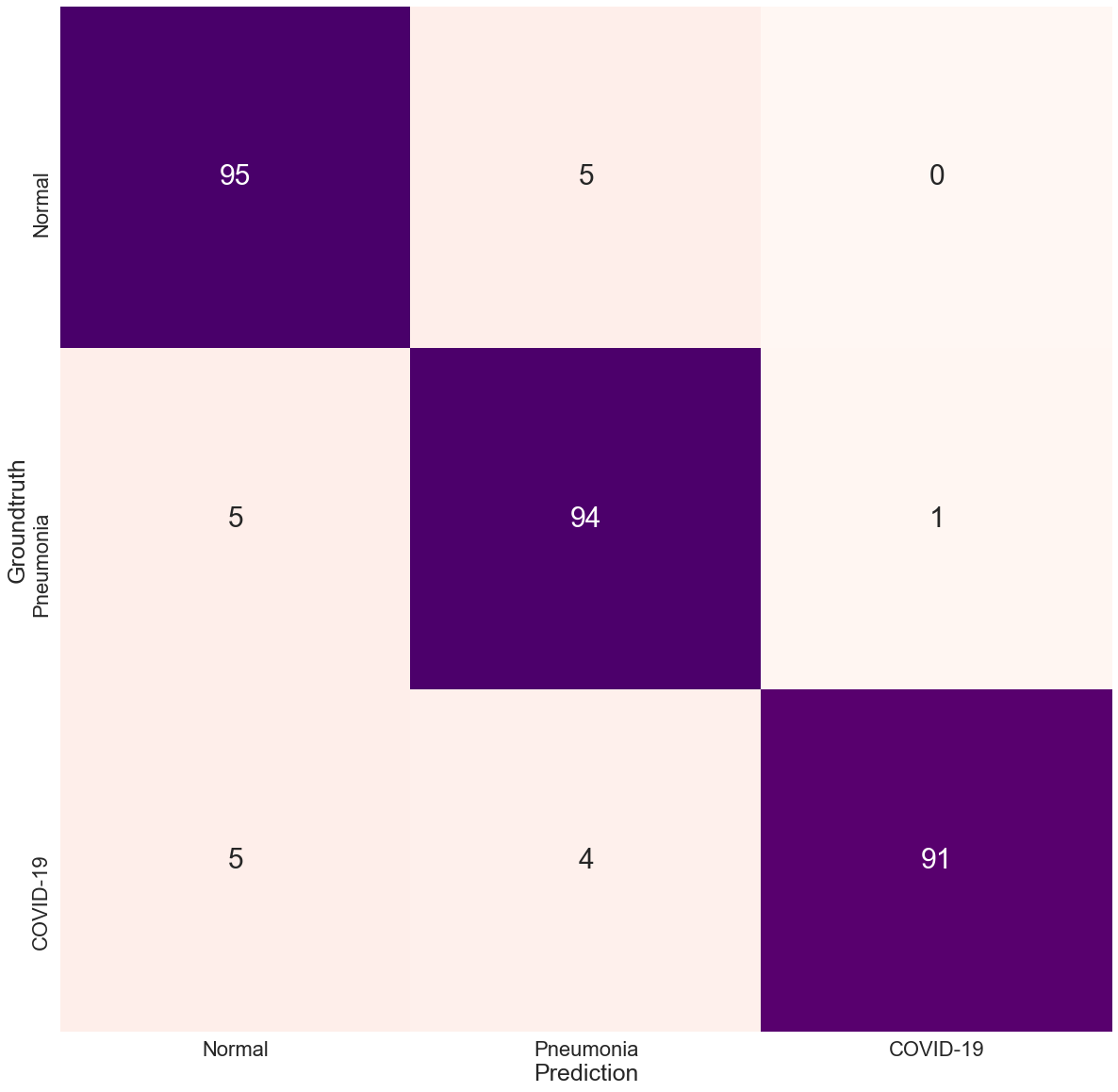}
  \caption{Confusion matrix for COVID-Net on the COVIDx test dataset.}
  \label{confusion}
\end{figure}

\begin{table}[h]
\caption{{Performance of tested deep neural network architectures on COVIDx test dataset. Best results highlighted in \textbf{bold}.}}
\begin{center}
\begin{tabular}{|c|c|c|c|}
\hline
 \textbf{Architecture} &  \textbf{Params (M)} &  \textbf{MACs (G)} & \textbf{Acc. (\%)}\\
\hline
VGG-19 & 20.37 & 89.63 & 83.0 \\
ResNet-50 & 24.97 & 17.75 & 90.6 \\
COVID-Net & \textbf{11.75} & \textbf{7.50} & \textbf{93.3} \\
\hline
\end{tabular}\par
\bigskip
\label{tab:static_comp_table}
\end{center}
\vspace{-20pt}
\end{table}

\begin{table}[h]
\centering
\caption{Sensitivity for each infection type. Best results highlighted in \textbf{bold}.}
\begin{adjustbox}{max width=\linewidth}
\begin{tabular}{|c|c|c|c|}
\hline
\multicolumn{4}{|c|}{\textbf{Sensitivity (\%)}} \\ \hline
\textbf{Architecture} & \textbf{Normal} & \textbf{Non-COVID19} & \textbf{COVID-19} \\ \hline
VGG-19 & \textbf{98.0} & 90.0 &  58.7 \\
ResNet-50 & 97.0 & 92.0 &  83.0 \\ 
COVID-Net & 95.0 & \textbf{94.0} &  \textbf{91.0} \\ \hline
\end{tabular}%
\end{adjustbox}
\label{tab:sensitivity}
\end{table}

\begin{table}[h]
\centering
\caption{Positive predictive value (PPV) for each infection type.  Best results highlighted in \textbf{bold}.}
\begin{adjustbox}{max width=\linewidth}
\begin{tabular}{|c|c|c|c|}
\hline
\multicolumn{4}{|c|}{\textbf{Positive Predictive Value (\%)}} \\ \hline
\textbf{Architecture} & \textbf{Normal} & \textbf{Non-COVID19} & \textbf{COVID-19} \\ \hline
VGG-19 & 83.1 & 75.0 & 98.4  \\ 
ResNet-50 & 88.2 & 86.8 & 98.8  \\ 
COVID-Net & \textbf{90.5} & \textbf{91.3} & \textbf{98.9}  \\ \hline
\end{tabular}%
\end{adjustbox}
\label{tab:ppv}
\end{table}

\begin{figure*}[h]
\centering
  \includegraphics[width=\textwidth]{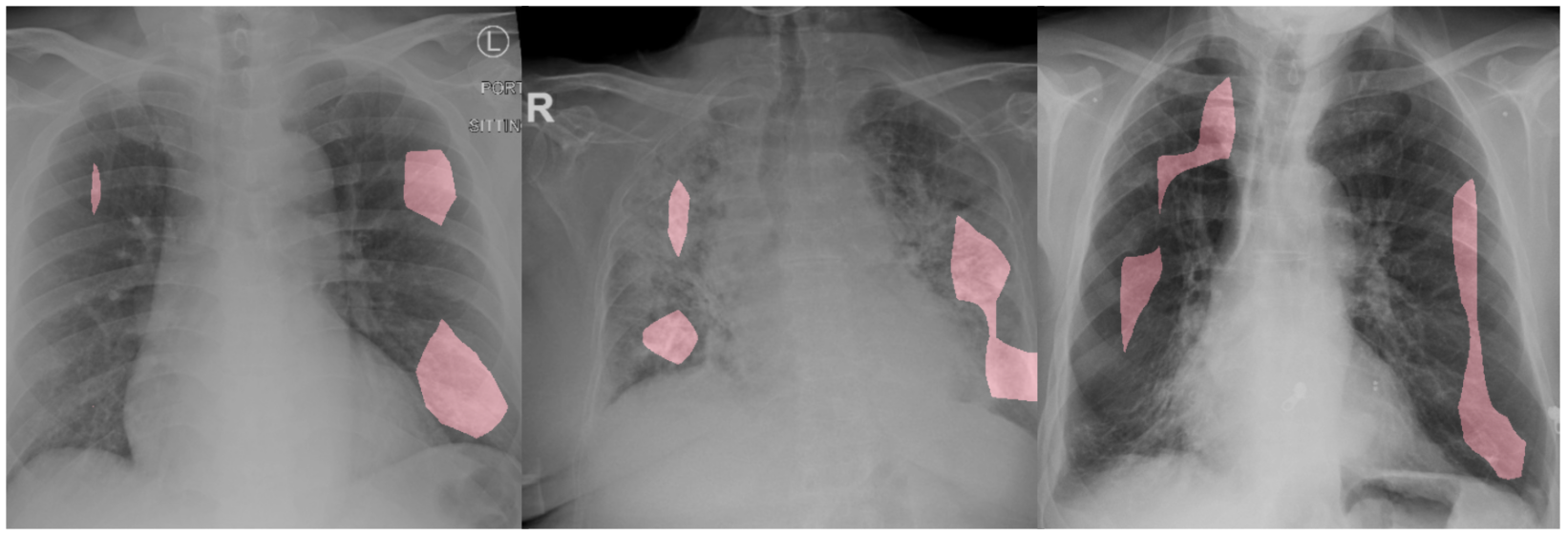}
  \caption{Example CXR images of COVID-19 cases from several different patients and their associated critical factors (highlighted in red) as identified by GSInquire~\cite{lin2019explaining}.}
  \label{examples}
\end{figure*}

 Next, we take a deeper exploration into the current limitations of the proposed COVID-Net by studying the sensitivity and PPV for each infection type, which is shown in Table \ref{tab:sensitivity} and Table \ref{tab:ppv}, respectively, and the confusion matrix in Fig. \ref{confusion}.  A number of interesting observations can be made about how COVID-Net performs under the different scenarios.  First, it can be observed that COVID-Net can achieve good sensitivity for COVID-19 cases (91.0\% sensitivity), which is important since we want to limit the number of missed COVID-19 cases as much as possible.  Second, it can be observed that COVID-Net achieves high PPV for COVID-19 cases (98.9\% PPV), which indicates very few false positive COVID-19 detections (for example, as seen in Fig. \ref{confusion}, one patient with non-COVID19 infection was misidentified as having COVID-19 viral infections). This high PPV is important given that too many false positives would increase the burden for the healthcare system due to the need for additional PCR testing and additional care.  Therefore, based on these results, it can be seen that while COVID-Net performs well as a whole in detecting COVID-19 cases from CXR images, there are several areas of improvement that can benefit from collecting additional data, as well as improving the underlying training methodology to generalize better across such scenarios.
 
 \subsection{Architecture Comparisons}
 We now take a deep exploration into the impact of architectural design choices made by generative synthesis during the machine-driven design exploration process on the resulting COVID-Net network architecture being able to achieve in terms of balance between computational efficiency and performance.  In order to perform this analysis, we evaluated the performances of the following deep neural network architectures for comparative purposes:
 \begin{itemize}
     \item \textbf{VGG-19~\cite{simonyan2014deep}}: A deep neural network architecture that does not leverage residual design principles, lightweight design patterns, and have very low architectural diversity.
     \item \textbf{ResNet-50~\cite{resnet}}: A deep neural network architecture that leverages residual design principles and lightweight design patterns (e.g., bottleneck design patterns) and have moderate architectural diversity, but does not leverage lightweight PEPX design patterns and selective long-range connectivity.
\end{itemize}     

The choice of these two deep neural network architectures is based on the fact that they do not possess the key defining traits of COVID-Net, which are lightweight PEPX design patterns, selective long-range connectivity, and high architectural diversity, thus allowing for a better understanding of the benefits of the unique traits of COVID-Net.

It can be observed in Table \ref{tab:static_comp_table} that COVID-Net had noticeably lower architectural complexity and computational complexity than the VGG-19 and ResNet-50 architectures.  For example, the proposed COVID-Net requires $\sim$12$\times$ fewer MAC operations than the VGG-19 architecture and $\sim$2.37$\times$ fewer MAC operations than the ReNset-50 architecture, respectively.  This illustrates the benefits of the lightweight PEPX design patterns within the COVID-Net architecture compared to not using lightweight design patterns (e.g., VGG-19) and using other types of lightweight design patterns (e.g., bottleneck patterns as used in ResNet-50).  It can also be observed in Table \ref{tab:static_comp_table}, \ref{tab:sensitivity}, and \ref{tab:ppv} that COVID-Net achieved noticeably higher test accuracy and COVID-19 sensitivity than the VGG-19 and ResNet-50 network architectures.  For example, COVID-19 sensitivity of COVID-Net is $>$32\% higher than VGG-19 and 8\% higher than ResNet-50.  These results illustrates the benefits of the selective long range connectivity and high architectural diversity found in COVID-Net, which enables strong representational capacity that is tailored for the task as well as making it easier to train.  Therefore, these results demonstrate the benefits of the different design choices made during the machine-driven design exploration stage of the human-machine collaborative design strategy employed to create COVID-Net.

\subsection{Qualitative Analysis} As mentioned earlier, we performed an audit on the proposed COVID-Net to gain better insights into how COVID-Net makes decisions, and validate whether it is making detection decisions based on relevant information rather than erroneous information that bias decisions based on irrelevant visual indicators.  The critical factors identified by GSInquire~\cite{lin2019explaining} in several example CXR images of COVID-19 cases are shown in Fig.~\ref{examples}.  It can be observed that, based on the interpretation produced by GSInquire, the proposed COVID-Net primarily leverages areas in the lungs in the CXR images as the main critical factors in determining whether a CXR image is of a patient with a SARS-CoV-2 viral infection, as shown in red in Fig.~\ref{examples}.  As such, we were able to validate that that COVID-Net was not relying on improper information to make decisions (e.g., erroneous visual indicators outside the body, embedded markup symbols, imaging artifacts, etc.), which could lead to scenarios where the right decisions are made for the wrong reasons.  Such `right decision, wrong reason' scenarios are very difficult to track and identify without the use of such an explainability-driven auditing strategy, and thus highlight the value of explainability in improving the reliability of deep neural networks for clinical applications. 

In addition to performance validation for more responsible and transparent design, the ability to interpret and gain insights into how the proposed COVID-Net detects COVID-19 infections is also important for a number of other reasons:
\begin{itemize}
\item \textbf{Transparency.} By understanding the critical factors being leveraged in COVID-19 case detection, the predictions made by the proposed COVID-Net become more transparent and trustworthy for clinicians to leverage during their screening process to aid them in making faster yet accurate assessments.
    \item \textbf{New insight discovery.} The critical factors leveraged by the proposed COVID-Net could potentially help clinicians discover new insights into the key visual indicators associated with SARS-CoV-2 viral infection, which they can then leverage to improve screening accuracy.
\end{itemize}

%% file: content/conclusion.tex
In this study, we introduced COVID-Net, a deep convolutional neural network design for the detection of COVID-19 cases from CXR images that is open source and available to the general public.  We also introduce COVIDx, an open access benchmark dataset that is comprised of 13,975 CXR images across 13,870 patient cases from five open access data repositories.  Moreover, we investigated how COVID-Net makes predictions using an explainability method in an attempt to gain deeper insights into critical factors associated with COVID cases, which can aid clinicians in improved screening as well as improve trust and transparency when leveraging COVID-Net for accelerated computer-aided screening.

By no means a production-ready solution, the hope is that the promising results achieved by COVID-Net on the COVIDx test dataset, along with the fact that it is available in open source format alongside the description on constructing the open source dataset, will lead it to be leveraged and build upon by both researchers and citizen data scientists alike to accelerate the development of highly accurate yet practical deep learning solutions for detecting COVID-19 cases from CXR images and accelerate treatment of those who need it the most.  Future directions include continuing to improve sensitivity and PPV to COVID-19 infections as new data is collected, as well as extend the proposed COVID-Net to risk stratification for survival analysis, predicting risk status of patients, and predicting hospitalization duration which would be useful for triaging, patient population management, and individualized care planning.